\documentclass[fleqn,twoside]{article}
\usepackage{espcrc2}
\usepackage{graphicx}

\usepackage{graphicx}
\usepackage[figuresright]{rotating}

\newcommand{\AmS}{{\protect\the\textfont2
  A\kern-.1667em\lower.5ex\hbox{M}\kern-.125emS}}

\title{Strong coupling analysis of diquark condensation\thanks{Talk given 
       by V. Laliena}\thanks{Work supported by INFN-CICyT collaboration and
       MCYT (Spain), grant FPA2000-1252. V.L. is a Ram\'on y Cajal fellow.}}

\author{V. Azcoiti\address[UNIZAR]{Departamento de F\'{\i}sica Te\'orica, 
        Universidad de Zaragoza, E-50009 Zaragoza (Spain)}, 
        G. Di Carlo\address[LNGS]{INFN, Laboratori Nazionali del Gran Sasso,
        67010 Assergi,(L'Aquila) (Italy)},
        A. Galante\addressmark[LNGS]\address{Dipartimento di Fisica 
        dell'Universit\`a di L'Aquila, 67100 L'Aquila (Italy)}
        and
        V. Laliena\addressmark[UNIZAR]
       }
       
\begin{document}

\begin{abstract}
The phenomenon of diquark condensation at non-zero baryon
density and zero temperature is analyzed in the strong coupling
limit of lattice QCD.
The results indicate that there is attraction in the
quark-quark channel also at strong coupling, and that the attraction is
more effective at high baryon density, but for infinite coupling it is
not enough to produce diquark condensation. It is argued that the
absence of diquark condensation is not a peculiarity of the strong coupling
limit, but persists at sufficiently large finite couplings.
\vspace{1pc}
\end{abstract}

\maketitle

A weak coupling analysis of QCD, both at the level of one gluon exchange
as well as using the instanton induced interaction, reveals that the 
quark-quark scattering amplitude is attractive in the color anti-triplet 
channel and repulsive in the symmetric color sextet channel.  
The existence of such attractive force between quarks in the color 
anti-triplet channel is also supported by experiments, since
hadron phenomenology is best explained by models that assume baryons
consiting of a quark-quark anti-triplet color bound state whose color
is neutralized by the remaining quark \cite{phen}.

In the deconfined phase, at high baryon densities and low temperatures, 
the effect may be more dramatic. If the weakly 
interacting quarks tend to form a Fermi surface, the quark-quark
attraction will render it unstable via the Cooper mechanism. 
Hence, the phenomenon of color superconductivity will take place:
a diquark condensate will appear in the ground state, a gap will be 
opened in the spectrum and the Fermi surface will be destroyed. 
Many exotic phenomena at high baryon density may happen as a
consequence of diquark condensation, as color flavor locking and
unlocking, crystalline color superconductivity, etc. \cite{colorsup}.

The diquark operator is not gauge invariant, and
therefore general arguments forbide diquark condensation in the naive
way: it must be understood not as the spontaneous breaking 
of the gauge symmetry, but as a kind of Higgs mechanism whose physical
consequences are the phenomena we name color superconductivity.

The arguments leading to the existence of diquark condensation at high 
baryon density are based on weak coupling analysis --although some non 
perturbative effects are taken into account via instantons--, and
thus are only valid at assymptotically large densities. In addition, the
study is performed using effective four quark theories which are not
gauge invariance, and then it is difficult to trace the fate of gauge 
invariance. It is of utmost importance to study diquark condensation
beyond the weak coupling region, because of the lack of reliable tools, since
Monte Carlo simulations are not feasible at high density and low temperature
due to the sign
problem --however, see \cite{sign}--. Hence, it is interesting to turn to the 
opposite regime, the strong coupling domain. Recent attempts to study
finite density QCD in the strong coupling limit using hamiltonian techniques
have been developed in \cite{strong}. Here, we present the results concerning
diquark condensation obtained in \cite{monos} by using path integral 
techniques.

Let us consider QCD --the gauge group is SU(3)-- at zero temperature and
finite baryon chemical potential, regularized on a euclidean 
four dimensional lattice with four flavor of staggered fermions.    
Let us define the meson and baryon fields
\begin{eqnarray}
M(x) &=& \bar\psi^a(x)\psi^a(x)\, , \\
B(x) &=&\frac{1}{6}\epsilon_{abc}\psi^a(x)\psi^b(x)\psi^c(x)\, , \\
\bar{B}(x) &=&\frac{1}{6}\epsilon_{abc}\bar\psi^a(x)\bar\psi^b(x)\bar\psi^c(x) 
\, ,
\end{eqnarray}
and the diquark field
\begin{eqnarray}
D^a(x) &=& \epsilon_{abc}\psi^b(x)\psi^c(x)\, , \label{diquark} \\
\bar{D}^a(x) &=& \epsilon_{abc}\bar\psi^b(x)\bar\psi^c(x) 
\label{adiquark} \, . 
\end{eqnarray}
where summation over repeated color indexes $a,b,c$ is understood. 
Note that $B(x)=1/6\psi^a(x)D^a(x)$ and 
$\bar{B}(x)=1/6\bar\psi^a(x)\bar{D}^a(x)$. 
At strong
coupling, the gauge field can be integrated out and what remains is a 
local action
for the fermions, which, by gauge invariance, depends only on $M(x)$,$B(x)$,
and $\bar{B}(x)$. The action is hence multilinear in the quark variables
$\psi^a(x)$ and $\bar\psi^a(x)$. 

The action can be converted in a bilinear in the quark fields at the expense
of introducing a real colorless bosonic field, $\varphi(x)$ and a color
triplet complex bosonic field, $\phi^a(x)$, through a series of 
Hubbard-Stratonovich transformations \cite{monos}. After integrating out the
quark field, the lattice grand canonical partition function is given by
an effective bosonic action which depends only on the fields $\varphi(x)$,
$\phi^a(x)$ and $\phi_a^\dagger(x)$, and which is invariant under 
SU(3) color gauge
transformations, so that it actually depends only on the gauge invariant 
combination $\phi_a^\dagger(x)\phi^a(x)$, and on $\varphi(x)$.

The field $\varphi(x)$ indeed represents the meson field $M(x)$ since, by
introducing suitable sources in the action, it
is easy to see that $\varphi(x)=\mathrm c M(x)$, where $\mathrm c$ is a 
constant. Thus, $\varphi(x)$ is an order parameter of chiral symmetry and
its expectation value is proportional to the chiral condensate. Analogously,
it can be seen that the field $\phi_a^\dagger(x)$ --which is a color 
anti-triplet-- is linked to the diquark field by
$D^a(x) = F(x)\,\phi^\dagger_a(x)$,
where $F(x)$ is a gauge invariant functional of $\varphi(x)$,
$\phi^a(x)$ and $\phi_a^\dagger(x)$. As expected, diquark condensation
takes place if the expectation value of $\phi^a(x)$ is non-zero. 

Thus, the field $\phi^a(x)$ is a kind of Higgs field
which is useful to study color superconductivity in the same way as the
theory of Ginzburg-Landau describes electric superconductivity. Here,
however, the effective action for $\phi^a(x)$ has been obtained exactly
from QCD, albeit in the strong coupling regime, and there is no
gauge field coupled to it.

To study the phase diagram of the theory we used the semi-classical 
approximation. The effective potential is given by the bosonic effective
action evaluated at constant fields $\varphi(x)=\bar\varphi$ and 
$|\phi(x)|^2=v^2$. Its absolute minimum 
describes the equilibrium state at a given temperature and chemical potential.
Other local minima, if present, are associated with metastable states.
In the chiral limit and at zero temperature we get the following
minima of the classical effective potential \cite{monos}:
\begin{equation}
\begin{array}{lll}
v=0\, , & \bar\varphi=\pm\left(\sqrt{33}-5\right)^{1/2} 
& \forall \; \mu \\
v=\sqrt[4]{5}\, , & \bar\varphi=0 & \forall \; \mu \\
v=0\, , & \bar\varphi=0 & \forall \; \mu > 0.4416 
\end{array}
\end{equation}
Note that the positions of the minima are independent of $\mu$.

There is a first order phase transition at $\mu_c\approx 1.557$. For
$\mu<\mu_c$ the --degenerate owing to chiral symmetry- absolute minima are 
at $v=0$ and $\bar\varphi=\pm(\sqrt{33}-5)^{1/2}$, corresponding to a phase
with chiral symmetry spontaneously broken. For $\mu>\mu_c$ the absolute
minimum is at $v=0$ and $\bar\varphi=0$; chiral symmetry is restored in 
this phase. Obviously, the transition is first order. For $\mu<\mu_c$
the baryon density is zero, and three for $\mu>\mu_c$. Three quarks
per lattice site is the maximum allowed by Pauli's principle, so that
the transition separates a phase of zero baryon density from a phase
saturated of quarks. The system at any intermediate density is 
thermodynamically unstable and splits into domains of zero and
saturated baryon density.
This behavior is an artifact of the strong coupling limit and
has been observed in other strong coupling analysis at zero
temperature and finite chemical potential.

\begin{figure}[t!]
\centerline{\includegraphics*[width=11pc,angle=90]{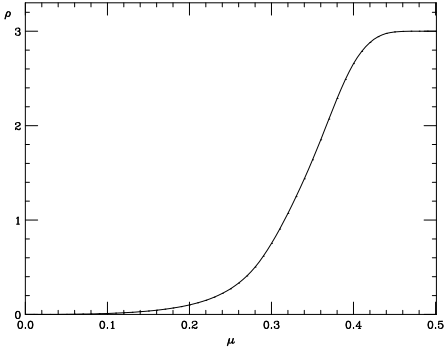}}
\vspace{-0.75truecm}
\caption{Baryon density as a function of the chemical potential in the
metastable state with diquark condensation.}
\label{fig:density}
\vspace{-0.5truecm}
\end{figure}

\begin{figure}[t!]
\centerline{\includegraphics*[width=11pc,angle=90]{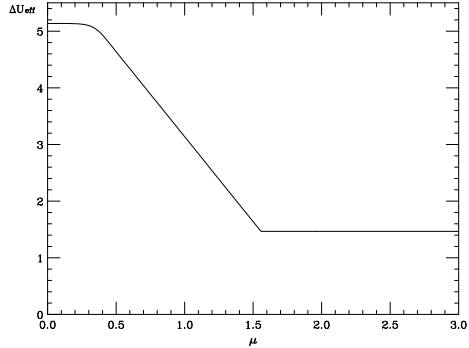}}
\vspace{-0.75truecm}
\caption{Energy difference between the diquark condensate metastable
state and the equilibrium state.}
\label{fig:barrier}
\vspace{-0.5truecm}
\end{figure}

The minimum at $v=\sqrt[4]{5}$ and $\bar\varphi=0$ is metastable in 
both phases. It describes a chiral symmetric state with a diquark 
condensate. The baryon density that correspond to this metastable
state is a smooth function of the chiral condensate 
(see figure~\ref{fig:density})
that interpolates between zero density and saturation.
This means that at any baryon density a state with diquark condensation
can be formed and have some short life until the system splits into
its zero density and saturated domains. The presence
of the metastable state signals the attraction in the quark-quark channel.
At strong coupling such attraction is not strong enough to form a stable 
diquark condensate. The energy difference between this metastable
state and the equilibrium state decreases with $\mu$,
indicating that the quark-quark attraction becomes the more effective the
higher the baryon density (see figure~\ref{fig:barrier}). However,
the energy difference is nonzero even at high $\mu$ (in the saturated 
phase). 

The crucial question is whether this 
metastable state becomes stable at sufficiently large density at finite 
coupling.
A partial answer is the following: 
assuming that the strong coupling expansion gives meaningful results,
diquark condensation cannot take place at sufficiently large couplings,
$g$, whatever the chemical potential. The reason is that, as we have seen,
the energy difference between the state with diquark condensation and 
the stable state remains positive for any value of the chemical potential 
in the strong coupling limit. A correction to the effective potential
of order $1/g$ cannot remove such energy difference if $g$ is 
sufficiently large. Hence, diquark condensation cannot take place
in the strong coupling region. If, on the other hand, diquarks condense
at some finite chemical potential in the weak coupling regime, there must 
be a critical value of the coupling, $g_c$, such that diquark condensation
takes place in some interval of the baryon density for $g<g_c$, but diquarks 
do not condense at any density for $g>g_c$. This means that the interval 
of baryon densities (in lattice units) at which diquark condensation occurs 
as a stable state shrinks as $g$ increases, and vanishes at $g_c$. 
For $g>g_c$ the state
with diquark condensate survives as a metastable state. 

\end{document}